\documentclass[aps,twocolumn,amsmath,amssymb,floatfix]{revtex4-1}

\usepackage{graphicx}
\usepackage{dcolumn} 
\usepackage{bm}      
\usepackage{color}
\usepackage{aas_macros}
\usepackage{ulem}

\renewcommand{\vec}{\mathbf}

\def\ADD#1{{#1}}           

\begin{document}

\title{Not Much Helicity is Needed to Drive Large Scale Dynamos}

\author{Jonathan Pietarila Graham$^{1}$, Eric G. Blackman$^{2}$, Pablo D. Mininni$^{3,4}$ and Annick Pouquet$^3$}
\affiliation{
$^1$Solid Mechanics and Fluid Dynamics (T-3) \& Center for Nonlinear Studies; Los Alamos National  Laboratory MS-B258; Los Alamos NM 87545, U.S.A. \\
$^2$Department of Physics and Astronomy; University of Rochester; Rochester NY 14627; U.S.A. \\
$^3$Computational and Information Systems Laboratory; NCAR; P.O. Box 3000, Boulder CO 80307-3000, U.S.A. \\
$^4$Departamento de F\'\i sica, Facultad de Ciencias Exactas y Naturales, Universidad de Buenos Aires \& IFIBA, CONICET; Ciudad Universitaria, 1428 Buenos Aires, Argentina.}
\date{\today}

\begin{abstract}
\ADD{Understanding the in situ amplification of large scale magnetic fields in turbulent astrophysical rotators has been a core subject  of dynamo theory.  
When turbulent velocities are helical, large scale dynamos that substantially amplify fields 
 on scales that  exceed the turbulent forcing scale arise, but the minimum sufficient fractional kinetic helicity $f_{h,C}$ has not been previously well quantified. 
Using direct numerical simulations for a simple helical dynamo, 
we show that
  $f_{h,C}$ decreases as the ratio of forcing to large scale
  wave numbers $k_F/k_{min}$ increases. From the condition that a large scale helical dynamo must overcome the backreaction  from any non-helical field on the large scales, we develop a theory  that can explain the simulations.
    For \ADD{$k_F/k_{min}\ge8$ we
    find $f_{h,C}\lesssim 3\%$,} 
  implying that very small
  helicity fractions strongly influence magnetic spectra for {even
    moderate}
scale separation.}
\end{abstract}
\maketitle

{\it Introduction}
\ADD {The  origin of  magnetic fields in  turbulent astrophysical rotators such as stars, galaxies   \cite{BeBrMo+1996Be2009}
and accretion disks  has been a long standing topic of research. A particular challenge has been to understand the origin of fields on scales that are large  compared to those of any underlying  turbulence
\cite{Lathrop2011,BeBrMo+1996Be2009,Br2009,blackman10}. }

\ADD{That the large scale  field of the sun reverses every 11 years
reveals that  such stellar fields cannot be simply the residual of flux freezing from the primordial material and must be amplified in situ.   Complementarily, the continuous  
processing by supernovae driven turbulence in galaxies  likely renders  the role of any primordial fields  to be  simply seed fields whose in situ processing must be understood  to account for the observed present  day large scale  fields in galaxies. 
The presence of astrophysical jets from accretion engines  also highlights the presence of large scale fields in accretion disks, and accretion disk simulations  \cite{Br2009} commonly show the in situ  generation of large scale magnetic fields that reverse on cycle periods of tens of orbit times.  }

\ADD{The study of  in situ field amplification in the presence of   velocity flows is the enterprise of dynamo theory.  Small scale dynamos (SSDs), in which turbulent velocity flows amplify fields at or below scales of the forcing  \cite{kazantsev,schek04},  can be distinguished from 
large scale dynamos (LSDs) in which magnetic fields are amplified on spatial or temporal variation scales larger than the scales of the underlying forcing. 
 LSDs and SSDs are often contemporaneous and interactive (see e.g. \cite{axel_rev,blackman10})
but   LSDs   arise only  when  turbulent velocities are  sufficiently helical \cite{PFL76,MeFrPo1981}.    There has been  little previous work, however, on  determining  the minimum sufficient helicity to  incite LSD action and this  is the topic of the present paper.
 Astrophysical flows are unlikely to be 100\% helical in environments where
 LSDs are presumed; the galaxy for example is estimated to have helicity of 
 $< 10\%$.  Thus    the basic question of how much helicity is required in even the simplest LSDs
 is important in assessing the potential ubiquity of LSDs.}


The standard 20th century textbook \cite{ Mo1978} kinematic approach to LSD theory has been 
classical mean field  (MFT)  which features the $\alpha-$effect:
  $\gamma = |\alpha|k - \beta k^2$, where $\gamma$ is the exponential
  growth rate in the kinematic dynamo regime (presuming that any Lorentz-force
  feedback is negligible), $k$ is the wavenumber of magnetic field
  growth, the $\alpha-$effect is proportional to kinetic helicity
  $H_v=\langle{\bf v} \cdot \boldsymbol{\omega}\rangle$, with ${\bf
    v}$ the velocity and ${\mathbf \omega}=\nabla \times {\bf v}$ the
  vorticity, and $\beta$ is the turbulent eddy diffusivity
  \cite{Mo1978}.  Such mean-field theory has been used to model
  solar \cite{BMT04Br2008}, stellar, and galactic observations, as well as   laboratory plasma dynamos \cite{GaLiPl+2002JiPr2002}, and  Geo-dynamos
\cite{RoGl2000Bu2000}.


But the kinematic approach to LSD theory is incomplete.
Although many astrophysical rotators have differential
rotation and open boundaries,  substantial progress in going beyond the kinematic theory has emerged from studies of the closed volume ``$\alpha^2$'' helical dynamo without
shear, in which the evolution of an initially weak seed field is
subject to helical velocity forcing.  The $\alpha^2$ dynamo was first
tackled semi-analytically \cite{PFL76} using a spectral
integro-differential model with an Eddy Damped Quasi-Normal Markovian
(EDQNM) closure, consistently tracking the magnetic helicity. It was
shown that the actual driver of large scale magnetic field growth is
not just the kinetic helicity, but the
\ADD{residual} helicity, $H_R=H_v-H_j$ \ADD{where} the current
helicity $H_j=\langle{\bf j} \cdot {\bf b}\rangle$ \ADD{and} ${\bf
  j}=\nabla \times {\bf b}$ is the current density.

This $\alpha^2$ dynamo in a periodic box was simulated
\cite{brandenburg01} by forcing with kinetic helicity at wavenumber
$k_F=5 k_{min}$ ($k_{min}=1$ was the smallest wavenumber of the
flow). The large-scale ($k<5$) field grew as expected from
Ref.~\cite{PFL76}. Subsequently, a two-scale $\alpha^2$ LSD was
developed \cite{Blackman02}; it incorporated magnetic helicity
evolution using a simpler closure than EDQNM and showed even a
two-scale nonlinear theory predicts the evolution and saturation of
 LSD growth observed in \cite{brandenburg01}. Driving with kinetic helicity initially
produces a large scale helical magnetic field, but the near
conservation of magnetic helicity leads to a compensating small scale
magnetic (and current) helicity of opposite sign. This counteracts the
kinetic helicity driving in the large scale field growth coefficient,
and quenches the LSD, as proposed in \cite{PFL76}.

\ADD{There has also been a plethora of work on the SSD.  In a periodic box
with a weak initial seed field and non-helical forcing, the stochastic
line stretching produces negligible field growth above the forcing
scale. Simulations of non-helical SSDs without large scale shear 
show  that the total magnetic 
energy is amplified to near equipartition with the total kinetic energy
not only in the kinematic regime as predicted
by \cite{kazantsev},
 but also in the saturated regime for large
magnetic Prandtl number $P_M=\nu/\eta$, where $\nu$ and $\eta$ are the
viscosity and magnetic diffusivity \cite{haugen0304, schek02,schek04},
as reviewed in \cite{axel_rev}. Astrophysical plasmas such as
the Galactic interstellar medium do not seem to exhibit this pile-up
\cite{minter96}.} 

To address this disparity between simulations of non-helical SSDs  and how conditions favorable for LSD growth might influence the magnetic spectrum on both large and small scales, results for dynamos in a periodic box forced with different amounts of fractional kinetic helicity $f_h$ (a dimensionless measure of the  degree of alignment between the velocity and the vorticity of the forcing function), were studied \cite{maron02}. It was found that  the magnetic spectrum above and below the forcing scale were contemporaneously affected by a sufficient $f_h$. The large scale field grew, and the magnetic spectrum at large wave numbers steepened. For $f_{h}=1$ and $f_h=0$, the results of \cite{brandenburg01} and \cite{schek02} were respectively recovered.

\ADD {But the restriction in \cite{maron02} to a forcing scale of $k_F=5$ and resolution of $64^3$ grid points left key unexplored questions. In particular, the minimum $f_h$ for LSD action, $f_{h,C}$, could not be determined as a function of $k_F$. The smaller this minimum,  the potentially more ubiquitous  LSD conditions  are in astrophysics. Here we perform much higher resolution simulations for fractionally helical dynamos and quantify how $f_{h,C}$ depends on $k_F/k_{min}$. We also develop a theory that correctly predicts the dependence,  seen in the simulations.}

{\it Equations and set-up}- 
The incompressible MHD equations for velocity $\vec{v}$ and magnetic field  $\vec{b}$ are:
\begin{eqnarray}
 \partial_t\vec{v} + \boldsymbol{\omega} \times \vec{v}
 = \vec{j} \times \vec{b} - \boldsymbol{\nabla} p + \nu \nabla^2 \vec{v} + \vec{F} \nonumber \\ 
{\partial_t \vec{A} =  \vec{v} \times \vec{b} -\nabla \phi + \eta \nabla^2 \vec{A}} \nonumber   \\
\boldsymbol{\nabla} \cdot \vec{v} =  0\,, \ \ \ {\boldsymbol{\nabla} \cdot \vec{A} = 0.}
\label{eq:mhd} \end{eqnarray}
The total pressure divided by the constant (unit) density $p$ and the potential $\phi$ are obtained self-consistently to ensure incompressibility and the Coulomb gauge. The Reynolds number is $Re=U_{rms}L_0/\nu$, with $U_{rms}$ and $L_0=2\pi\int E(k)k^{-1}dk/\int E(k)dk$ the r.m.s.~velocity and the integral scale respectively; the magnetic Reynolds number is defined as $R_M=U_{rms}L_0/\eta$. In the following, $E$ denotes the total energy, and $E_v$ and $E_b$ denote the kinetic and magnetic energy respectively.


We employ a well-tested pseudo-spectral code that uses a hybrid parallelization, combining Message Passing Interface (MPI) and OpenMP \cite{mininni_hybrid}. The computational box  has size $[2\pi]^3$, and wave numbers vary from $k_{min}=1$ to $k_{max}=N/3$ using a standard 2/3 de-aliasing rule, where $N$ is the number of grid points per direction.

\begin{figure}[htbp]
\includegraphics[width=8.8cm]{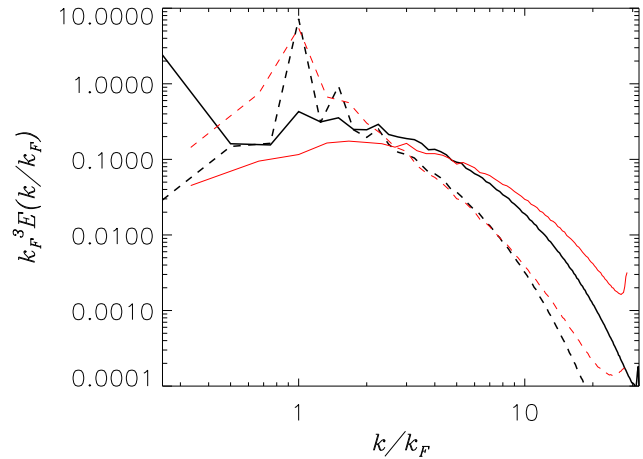}
\caption{({\it Color online}) Re-dimensionalized magnetic (solid) and
  kinetic (dashed) energy spectra after 90$\tau$ for run 4-60
  ({thick) black} forced at $k_F=4$ and for run 3-60 (thin red/light
  gray) with $k_F=3$. At a fixed $f_h=60\%$ here, increased scale
  separation provides for the transition between SSD and LSD.}
\label{fig:fig_spec}
\end{figure}

The forcing  applied at $k_F$ is
$\vec{F}\equiv\vec{F}_R+c\vec{F}_A$; $\vec{F}_A$ is an ABC flow
at $k_F$, and $\vec{F}_R$ is the sum of all harmonic
modes with {$k=k_F$} and random phases. We choose
$c$ for a given
fractional helicity of $\vec{F}$, $|f_h| \le 1$, with $f_h \equiv
\langle\vec{F}\cdot\boldsymbol{\omega_F}\rangle
(\langle|\vec{F}|^2\rangle \langle|\boldsymbol{\omega_F}|^2\rangle
)^{-1/2}$, where ${\mathbf \omega}_F{=\nabla\times\vec{F}}$. The entire forcing has random phases applied, with a correlation time $t_{cor}=0.1$. In practice,
the ratio of helical to non-helical forcing magnitudes is $c\simeq
[f_h/(1-f_h)]^{1/2}\equiv R_h$.  \ADD{The kinetic helicity
  at $k_F$ is typically within
  25\% of $f_h$.}  By choosing dimensional length and time constants
$l_0\propto k_F$ and $t_0$ {(fixed),} varying $k_F$ in our
simulations corresponds to dimensionalized
physical systems described by Eqs.~(\ref{eq:mhd}), where the
forcing scale is constant and the system size increases $\propto k_F$. The
dimensionless velocity $v_{rms}$ and forcing are $\propto
k_F^{-1}$, and the diffusivity $\propto
k_F^{-2}$.

A hydrodynamic state is evolved for five
\ADD{forcing-scale eddy}
 turnover times, $\tau$, before a magnetic seed field at
$k=k_{seed}$ is introduced. In the hydro steady state, the
resulting  $\tau=2\pi [k_F
  U_{rms}]^{-1} \approx 4.2$. In all simulations, dimensionalized
viscosity is constant, $k_F^2\nu=2.412\cdot10^{-2}$, \ADD{and,
  arbitrarily, $P_M=4$ so that $P_M>1$ while limiting the
  computational cost} (see Table \ref{tab1} for further details).

\begin{table*}
\caption{Parameters: Runs  are labeled by the \ADD{forcing wavenumber $k_F$}
  followed by the percentage of helicity in the
  forcing $f_h$; $R_M$ and $k_{seed}$ are defined in the text
  ($Re=R_M/4$). The SSD growth rate is $\gamma_{SSD}$. $E_b^{s}$ is
  the magnetic energy \ADD{and}
  $H_b$ is the magnetic helicity, \ADD{both} at later times, between \ADD{$60\tau$ and
  $130\tau$;}  the ``$f$''
  is for fluctuating, and ``$g$'' is for growing exponentially (indicating a helical dynamo).  Forcing wave numbers are
  $k_F=2,3,4,5,6$ \ADD{ and $8$} for runs on grids of $192^3,256^3,384^3,432^3,512^3$\ADD{
  and $ 768^3$} points respectively. ${\bf ^{\ast}}$Run 3-80 was
  pursued until $t\approx 300\tau$, at which time $E_b\sim \ADD{0.3}$ and
  $H_b\sim 10$. \ADD{See Fig. \ref{fig:find_growth}.}}
\begin{ruledtabular}
\begin{tabular}{ccccccc|ccccccccc}
Run    & $R_M$ & $k_{seed}$  & $\gamma_{SSD}$ & $\gamma_{k=1}$        & $E_b^s$ & $-100H_b$ &
Run    & $R_M$ & $k_{seed}$  & $\gamma_{SSD}$ & $\gamma_{k=1}$        & $E_b^s$ & $-100H_b$ \\
\hline
2-80 & 1500   &   [6.7,10.7]   &    0.24        & \ADD{$(-1.2\pm10)10^{-4}$} & 0.2    &   $0.5f$  &
5-09 & 2000   & [16.7,26.7] &    0.26        & \ADD{$(5.9\pm4.5)10^{-4}$} & $0.03$ & \ADD{$0.004f$} \\

\ADD{2-85} & 1600   &   --   &    0.22        & {$(5.6\pm0.7)10^{-3}$} & 0.4   &   $6g$  &
\ADD{5-19} & 1900   & -- &    0.26        & $(-2.5\pm0.7)10^{-3}$           & 0.03    &   $0.007f$\\

2-90 & 1600   &   --   &    0.24        & \ADD{$(6.0\pm0.7)10^{-3}$} & 0.4    &   $8g$  &
\ADD{5-40} & 1800   & -- &    0.27        & $(1.6\pm0.7)10^{-3}$           & 0.03    &   $0.03g$\\

\cline{1-7}
3-40 & 2000   &   [10,16]   &    0.31        & $(-1.0\pm7.6)10^{-4}$ & 0.1     &   $0.1f$  &
\ADD{5-50} & 1900   & -- &    0.27        & $(3.5\pm1.4)10^{-3}$           & 0.03    &   $0.06g$\\

3-60 & 1900   &     --      &    0.32        & $(3.7\pm7.3)10^{-4}$  & 0.1     &   $0.2f$  &
\ADD{5-60} & 1800   & -- &    0.28        & $(1.1\pm0.09)10^{-2}$           & 0.04    &   $0.3g$\\

\cline{8-14}
3-69 & 1700   &     --      &    0.27        & $(6.0\pm0.7)10^{-3}$  & 0.1     &   $1.0g$  &
{6-01} & 1700   &    [20,32]     &    0.27        & \ADD{$(2.2\pm11)10^{-4}$}  & 0.02    & $0.002f$ \\

3-80 & 2000   &     --      &    0.33        & $(8.6\pm1.4)10^{-3}$  & $0.3^{\ast}$ & $10g^{\ast}$ &
{6-05} & 1700   &     --      &    0.24        & \ADD{$(4.1\pm5.0)10^{-4}$}  & 0.02    & $0.003g$ \\

\cline{1-7}
4-10 & 1700   & [13.3,21.3] &    0.25        & $(-1.6\pm2.0)10^{-3}$ & 0.04    &  $0.008f$ &
\ADD{6-10} & 1700 & -- & 0.27 & $(-1.5\pm6.9)10^{-4}$ & 0.02 & $0.004g$\\

4-20 & 1600   &     --      &    0.28        & $(5.9\pm0.6)10^{-3}$  & 0.04    &  $0.06g$ &
\ADD{6-15} & 1600 & -- & 0.23 & $(1.1\pm0.7)10^{-3}$ & 0.02 & $0.007g$\\

4-40 & 1600   &     --      &    0.25        & $(1.5\pm0.1)10^{-2}$  &  0.06    &   $0.3g$  &
\ADD{6-20} & 1700 & -- & 0.27 & $(4.5\pm2.0)10^{-3}$ & 0.03 & $0.01g$\\

4-60 & 1500   &     --      &    0.25        & $(2.8\pm0.2)10^{-2}$  & 0.1     &   $1.0g$   &
\ADD{6-30} & 1600 & -- & 0.23 & $(3.6\pm0.9)10^{-3}$ &0.03  & $0.02g$ \\

4-80 & 1600   &     --      &    0.27        & $(2.8\pm0.3)10^{-2}$  & 0.1     &   $1.9g$     &
\ADD{6-40} & 1600 & -- & 0.23 & $(7.4\pm0.7)10^{-3}$ & 0.03  & $0.06g$ \\

\cline{1-7}
\cline{8-14}
&    &        &      &   &      &       &
\ADD{8-03} & 1200 & [26.7,42.7] & 0.20 & $(5.4\pm1.9)10^{-3}$ & 0.008 & $4\cdot10^{-4}g$ \\
\end{tabular} \end{ruledtabular}
\label{tab1} \end{table*}

{\it Simulation  Results}- 
Table \ref{tab1} summarizes our runs. The kinetic and magnetic energy spectra after $t= 90\tau$ are displayed in Fig.~\ref{fig:fig_spec}, for runs 4-60 (with $k_F=4$) and 3-60 (with $k_F=3$). Both have $f_h$=60\%, and for both, the small-scale fields grow; only for $k_F=4$   does $k=1$ grow.

\begin{figure}
\includegraphics[width=8.8cm]{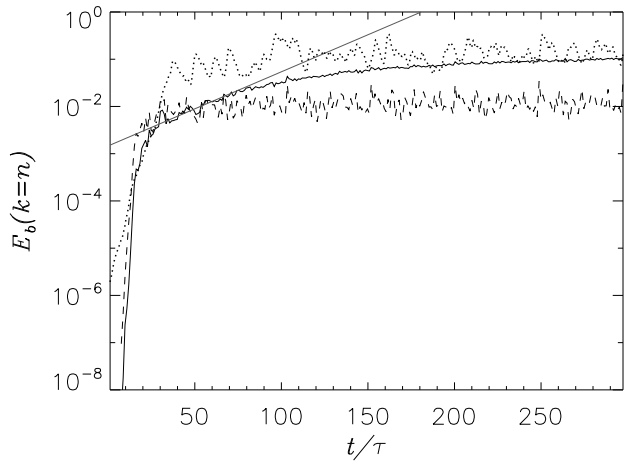}
\caption{Magnetic energy density $E_b(k)$ for $k=1$ (solid line),
  \ADD{$k=6$ (dashed) and total (dotted) versus time for run 3-80. The gray line
    indicates the fit {$\gamma=(8.6\pm1.4)10^{-3}$} to the solid curve
    for growth of the $k=1$ mode after SSD saturation. }}
\label{fig:find_growth}
\end{figure}

\begin{figure}
\includegraphics[width=8.8cm]{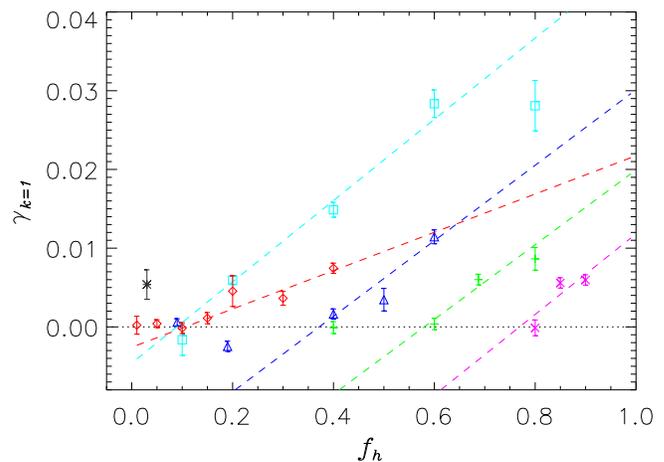}
\caption{Growth rate $\gamma_{k=1}$ of $E_b(k=1)$
  versus fractional helicity.  From Table \ref{tab1}, $\times$ (pink, $k_F=2$), + (green, $k_F=3$), $ \square$ (cyan, $k_F=4$),
  $\triangle$ (blue, $k_F=5$), $\diamond$ (red, $k_F=6$), \ADD{grey, $\ast$
    ($k_F=8$) and least-squares linear
    fits (dashed).}}
\label{fig:plot_growth}
\end{figure}

Figure \ref{fig:find_growth} shows the growth of magnetic energy in
the $k=1$, \ADD{$k=6$, and total over all modes for run 3-80.} The
evolution exhibits an early phase in which both modes grow at the same
rapid rate, with $\gamma_{SSD}\sim \ADD{0.33}\sim \tau^{-1}$, followed by a
slow growth of the $k=1$ mode and a saturation for the \ADD{$k=6$} mode.
\ADD{The $k=1$ mode accounts for nearly 10\% of $E_b$ by $100 \tau$
  and the growth rate slows, but  has not  fully saturated by
  $300 \tau$.}  The growth rate of magnetic energy at $k=1$ during the
SSD phase is nearly the same for all of our runs, and is
insensitive to $f_h$, $R_M$, and $k_F$. Sensitivity to $f_h$ emerges
once the SSD regime ends.
 
The $\gamma_{k=1}$ growth rates (for $E_b(k=1)$) that immediately
follow the SSD phase (see Table \ref{tab1}) are shown in
Fig. \ref{fig:plot_growth}. This LSD growth regime occurs only when $f_h> f_{h,C}$;
the LSD growth rate varies {linearly with $f_h$ for a fixed
  $k_F/k_{min}$} \ADD{($\alpha\propto H_v \propto{f_h}$).
  Least-squares fits are dashed lines in Fig. \ref{fig:plot_growth}
  (the y-intercept, $\beta\sim (k_{min}/k_F)^{2.2}$).  The
  short exponential growth phase of 4-80 makes for an
  inaccurate measure of $\gamma_{k=1}$; it is thus excluded
  from the fit.} As $k_F/k_{min}$ increases, 
 $f_{h,C}$ decreases. 
 
\begin{figure}
\includegraphics[width=8.8cm]{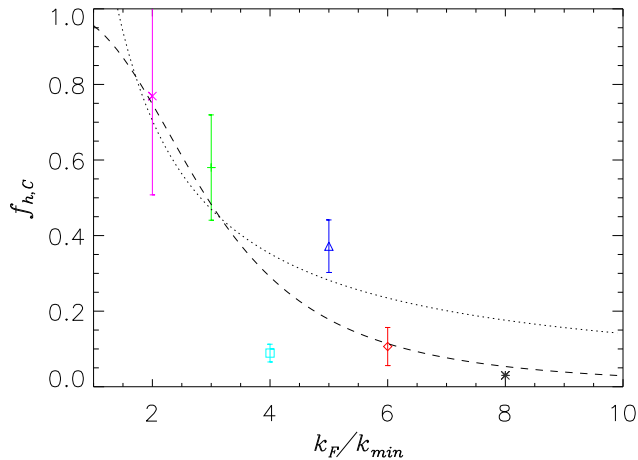}
\caption{$f_{h,C}$ \ADD{from least-squares fits}
  versus $k_F/k_{min}$ (symbols as in
  Fig.~\ref{fig:plot_growth}). Dashed line is best fit to
  Eq.~(\ref{eq:predict}), giving \ADD{$C=0.21$ and $\xi=0.46$.  Dotted
    line is kinematic MFT prediction $f_{h,C}=\beta k/(|\alpha_0| k_F)$. }}
\label{fig:plot_growthe}
\end{figure}

 The LSD exponential growth of $E_b(k=1)$ for $f_h \ge
f_{h,C}$ is accompanied by a $k=1$ growth of magnetic helicity. Studies of the $k=1$ growth for $f_h=1$ in a two-scale
approach \cite{brandenburg01,Blackman02} for a $H_R$
driven dynamo \cite{PFL76} suggest two phases of $k_{min}=1$ mode
growth after the SSD regime: one phase that is largely independent of
$R_M$, and a subsequent $R_M$ dependent asymptotic regime. The former
phase has growth consistent with our $\gamma_{k=1}$ phase.
\ADD{In} all these runs, $f_{h,C}$  decreases with increasing $k_{F}/k_{min}$, as displayed in Fig.~\ref{fig:plot_growthe}.  \ADD{For the largest $k_{F}/k_{min}$ (=8) case, $f_{h} \sim 3\%$ is sufficient.}
In Fig. \ref{fig:plot_growthe}, the error bars and $f_{h,C}$  are calculated as follows:
\ADD{The x-intercept from the least-squares linear fits, $\gamma_{k=1}=m f_h + b$ shown in
Fig.~\ref{fig:plot_growth}, determine our estimate of $f_{h,C}=-b/m$ for
Fig.~\ref{fig:plot_growthe}.  The 1$-\sigma$ uncertainties for $b$ and $m$
are then propagated for $f_{h,C}: \sigma_{f_{h,C}} = b/m
\sqrt{(\sigma_b/b)^2 + (\sigma_m/m)^2}$.}
 
{\it Theoretical prediction for $f_{h,C}$}- The following prediction
for $f_{h,C}$ is based on the principle that  LSD helical field
growth at $k=k_{min}$ beyond the SSD phase requires helical velocity forcing to overcome
the Lorentz force at $k=k_{min}$ at the end of the kinematic SSD
phase. For that time, we assume the magnetic energy at $k_{min}<k_F$
to be $B_{min}^2 \sim B_F^2 (k_{min}/k_F)^{\xi}$, where $B_F$ is the
magnetic field at $k_F$, and \ADD{$\xi-1$ is the slope of the magnetic
  energy spectrum on a log-log plot.} The associated Lorentz force is then $M_{nh} (f_h)
B_F^2 k_F (k_{min}/k_F)^{\xi+1}$, where the function $M_{nh}(f_h)<1 $
accounts for the contribution from only non-helical magnetic energy.

The available helical velocity forcing that must overcome this Lorentz
force is only a fraction of the helical forcing at $k=k_F$: At early
times when magnetic helicity is nearly conserved, the forcing not only sources
magnetic helicity at $k=k_{min}$ but also an oppositely signed, equal in
magnitude, magnetic helicity at $k=k_{ss}\ge k_F$. The associated
ratio of helical magnetic energy growth at $k_{min}$ to that at
$k_{ss}$ is then $ \sim k_{min}/k_{ss}<1$. The helical force that
needs to exceed the Lorentz force at $k_{min}$ to initiate growth is
thus $\sim K_h(f_h) v_{rms}^2 k_F(k_{min}/k_{ss})$, where the function
$K_h(f_h)<1$ accounts for only kinetic helical forcing.

Balancing the aforementioned forces assuming $K_{h}/M_{nh}= R_h(f_h)$ (which is consistent with our data), and assuming $f_h=f_{h,C}$, then {gives
\begin{equation}
f_{h,C}= { 1 \over 1 + C^2 (k_F/k_{min})^{2\xi+2}} ,
\label{eq:predict}
\end{equation}
where} $C\equiv (k_{min}v_{rms}^2)/(k_{ss}B_F^2) \sim
k_{min}/k_{ss}$. Figure \ref{fig:plot_growthe} shows the 
data and {the best fit using Eq.~(\ref{eq:predict}); 
  \ADD{$\xi=0.46\approx1/2$} is found.  This
  yields the prediction, $f_{h,C}\sim (k_F/k_{min})^\ADD{-3}$
  as $k_F/k_{min}\rightarrow\infty$.} {Note that taking
  the limit of infinite scale separation, we have a LSD with zero
  helicity (but only fluctuations, as in \cite{gilbert}).}

\ADD{The theory above, which considers the Lorentz force backreaction from the large scale field, can be contrasted with the prediction from the  purely kinematic theory of the standard $\alpha^2$ dynamo which does not include any  Lorentz forces. 
Using the formula presented in the introduction for the growth rate at $k_{min}$,  
and the definition of $f_h$, the critical fractional helicity for the kinematic theory would be} \ADD{$f_{h,C}=\beta k_{min}/(|\alpha_0| k_{f})$ where $\alpha_0\equiv\alpha/f_h$.}  \ADD{This formula is shown as the dotted line in Fig. \ref{fig:plot_growthe}
and does not fit the data very well, highlighting the importance of including the Lorentz force. 
This does not imply the kinematic theory is irrelevant however.
For values of $f_h >> f_{h,C}$ the kinematic theory should be  applicable to estimating the early time growth rate because the driving
helicity overwhelms the backreaction associated with the weak large scale field produced by the SSD
in that regime. }

 
{\it Discussion of LSD growth and saturation}-\ADD{At large $R_M$, SSD action produces field at all
  scales, potentially precluding a scale separation between the mean
  magnetic field and velocity fluctuations (an essential assumption to
  derive $\alpha^2$ MFT) \cite{CaHu2009}.  The SSD magnetic energy spectrum at scales above
    the forcing scale produces less magnetic energy   the larger the scale  \cite{kazantsev}.
    At large enough scales, the magnetic energy production from  the SSD  will be negligible, and scale separation  becomes a meaningful
  concept (see thin red/light gray, solid line in Fig. \ref{fig:fig_spec}).  This helps justify the mean field approach
  to LSDs.
}

\ADD{In the mean field, two-scale approach, once the small scale magnetic helicity has grown as a result of magnetic helicity conservation  to be large enough such that  the associated small scale current helicity  backreacts on the driving kinetic helicity,   the $\alpha^2$ dynamo eventually slows to $R_M-$dependent  growth rates and ultimately saturates completely.
 Previous studies   have typically  focused on the $f_h=1$ case \cite{brandenburg01,Blackman02}.
 In this paper, we have not run enough simulations long enough to determine how
strong the large scale field gets before its  evolution reaches the $R_M$ dependent regime.
 However, if cases with fractional helicity 
  $f_{h,C}< f_h < 1$ saturate by direct analogy to the $f_h=1$ cases studied in previous work, then the
 value of the large scale magnetic energy reached
just before the $R_M$ dependent regime emerges  would be expected to be simply proportional to an extra factor of   $f_h$, namely  ${\overline B}^2 \sim f_h (k_1 /k_f)\langle U_{rms}^2 \rangle$. Similarly,  for asymptotically saturated steady state at very late times, we would expect  ${\overline B}^2 \sim f_h (k_f /k_1)\langle U_{rms}^2 \rangle$. Note that in the $f_h=1$ case, the latter similarity highlights the fact that 
that  super-equipartition field strengths (with respect to the total kinetic energy) 
are able to  grow by the end of the nonlinear, saturated regime for fully helical $\alpha^2$ dynamo. }
 
Note however that the $R_M$ dependent regimes of the $\alpha^2$ dynamo are largely irrelevant for astrophysical objects which have such large $R_M$ that something else probably happens before these regimes are reached.
 Open boundaries and helicity fluxes are ingredients that have to be considered in realistic systems.   
In addition, real astrophysical dynamos have large scale shear, which amplifies the total large scale field beyond its purely helical value.  More work is needed to determine the strength of
the large scale fields produced by fractionally helical LSDs.
 
{\it Conclusion}- Only a minuscule amount of fractional helicity is
required for LSD action at even modest astrophysically relevant scale
separations.  For $f_h> f_{h,C}$,  the $k=1$ field grows and
 the small scale spectrum steepens (see Fig.~\ref{fig:fig_spec} and
{\cite{maron02}).  This {may be}  important because our result that
  \ADD{$f_{h,C}\lesssim 3\%$ for $k_F/k_{min}\ge8$,}  offers a basic principle
for  potentially reconciling a disparity between a pile up of small scale
  magnetic energy in large $P_M$ non-helical dynamo simulations
  \cite{schek04} and Galactic observations \cite{minter96}. 
  \ADD{Our results also suggest that 
   large scale separations should be a priority in designing  laboratory experiments to measure
    LSD action    \cite{Lathrop2011}.}

{\it Acknowledgements-} Computer time was provided by NCAR, Los Alamos
National Laboratory and Max-Planck-Institut f\"ur
Sonnensystemforschung. NCAR is sponsored by NSF. PDM was supported by
grants PIP 11220090100825 and PICT-2007-02211.  JPG gratefully
acknowledges the support of the U.S. Department of Energy through the
LANL/LDRD Program for this work.

%

\end{document}